\documentclass[manuscript]{aastex631}
\usepackage{amsmath}
\usepackage[section]{placeins}
\usepackage [english]{babel}
\usepackage [autostyle, english = american]{csquotes}
\MakeOuterQuote{"}
\usepackage{placeins}

\begin{document}


\title{The Independence of Magnetic Turbulent Power Spectra to the Presence of Switchbacks in the Inner Heliosphere}

\correspondingauthor{Peter D. Tatum}
\email{peter.tatum@lasp.colorado.edu}



\author[0000-0002-4626-4656]{Peter D. Tatum}
\affiliation{Laboratory for Atmospheric and Space Physics, University of Colorado, 1234 Innovation Drive, Boulder, CO 80303, USA}
\affiliation{Astrophysical and Planetary Sciences Department, University of Colorado, 80309 Duane Physics Building, Boulder, CO 80309, USA}

\author[0000-0003-1191-1558]{David M. Malaspina}
\affiliation{Laboratory for Atmospheric and Space Physics, University of Colorado, 1234 Innovation Drive, Boulder, CO 80303, USA}
\affiliation{Astrophysical and Planetary Sciences Department, University of Colorado, 80309 Duane Physics Building, Boulder, CO 80309, USA}

\author[0000-0001-8478-5797]{Alexandros Chasapis}
\affiliation{Laboratory for Atmospheric and Space Physics, University of Colorado, 1234 Innovation Drive, Boulder, CO 80303, USA}

\author[0000-0003-3945-6577]{Benjamin Short}
\affiliation{Laboratory for Atmospheric and Space Physics, University of Colorado, 1234 Innovation Drive, Boulder, CO 80303, USA}
\affiliation{Physics Department, University of Colorado, 80309 Duane Physics Building, Boulder, CO 80309, USA}



\begin{abstract}

An outstanding gap in our knowledge of the solar wind is the relationship between switchbacks and solar wind turbulence. Switchbacks are large fluctuations, even reversals, of the background magnetic field embedded in the solar wind flow. It has been proposed that switchbacks may form as a product of turbulence and decay via coupling with the turbulent cascade. In this work, we examine how properties of solar wind magnetic field turbulence vary in the presence or absence of switchbacks. Specifically, we use in-situ particle and fields measurements from Parker Solar Probe to measure magnetic field turbulent wave power, separately in the inertial and kinetic ranges, as a function of switchback magnetic deflection angle. We demonstrate that the angle between the background magnetic field and the solar wind velocity in the spacecraft frame ($\theta_{vB}$) strongly determines whether Parker Solar Probe samples wave power parallel or perpendicular to the background magnetic field. Further, we show that $\theta_{vB}$ is strongly modulated by the switchback magnetic deflection angle. In this analysis, we demonstrate that switchback deflection angle does not correspond to any significant increase in wave power in either the inertial range or at kinetic scales. This result implies that switchbacks do not strongly couple to the turbulent cascade in the inertial or kinetic ranges via turbulent wave-particle interactions. Therefore, we do not expect switchbacks to contribute significantly to solar wind heating through this type of energy conversion pathway, although contributions via other mechanisms, such as magnetic reconnection may still be significant.

\end{abstract}


\section{Introduction} 

Observations by Parker Solar Probe (PSP) have shown the ubiquity of switchbacks in the near-Sun solar wind \citep{bale_highly_2019,kasper_alfvenic_2019,dudok_de_wit_switchbacks_2020,horbury_sharp_2020}. Yet, the origin and evolution of switchbacks is still not well understood. Switchbacks remain a topic of interest because of their potential to elucidate processes taking place in the corona and the solar wind that may contribute to solar wind heating and acceleration. 

Switchbacks are characterized as large amplitude Alfvénic fluctuations, or even reversals, of the background magnetic field accompanied by radial velocity jets \citep{matteini_dependence_2014}. Recent work toward understanding the local properties of switchbacks has shown that they are three dimensional Alfvénic structures that propagate away from the sun \citep{horbury_sharp_2020,laker_statistical_2021,woodham_enhanced_2021}. One idea for the generation of switchbacks is that they are the result of interchange reconnection in the solar corona \citep{bale_solar_2021,laker_switchback_2022, wyper_imprint_2022, bale_interchange_2023,kumar_new_2023}. \cite{tenerani_magnetic_2020} used MHD simulations to show that the resulting large amplitude Alfvénic fluctuations can persist out to many solar radii where they can be detected by spacecraft, despite eventually decaying due to parametric decay instability. Another possibility for the generation of switchbacks is that they form in-situ throughout the solar wind and persist as the result of a balance between their generation and their decay \citep{schwadron_switchbacks_2021,tenerani_evolution_2021}. In particular, it has been proposed that switchbacks could form dynamically as a result of turbulent fluctuations propagating in the expanding solar wind \citep{squire_-situ_2020,shoda_turbulent_2021}. Whether or not switchbacks play a dynamical role in the solar wind turbulent cascade is of interest due to the significance of turbulence as a ubiquitous and uniform source of heating in the solar wind.

Observations throughout the heliosphere have shown that turbulence is ubiquitous in the solar wind, and plays an important role in its dynamics via turbulent dissipation and plasma energization \citep{verscharen_multi-scale_2019,bruno_solar_2013}. Previous observations have established the formation of an inertial range of shear Alfvén fluctuations that transitions to a kinetic range, often characterized by the presence of waves at proton scales, such as Kinetic Alfvén Waves (KAWs) and Ion-Cyclotron Waves, as well as intermittent structures such as reconnecting current sheets\citep{goldstein_magnetohydrodynamic_1999,salem_identification_2012,goldstein_kinetic_2015,bowen_constraining_2020}. Multiple pathways exist for kinetic-scale turbulent fluctuations to heat and accelerate solar wind plasma \citep{kletzing_electron_1994,leamon_dissipation_1999,chandran_perpendicular_2010,cranmer_proton_2012,klein_diagnosing_2017,isenberg_perpendicular_2019,klein_diagnosing_2020,bowen_constraining_2020}. However, which of those mechanisms control turbulent energization within solar wind turbulence has yet to be determined. If switchbacks affect the evolution of the turbulent cascade, their spatial inhomogeneity in the near-Sun solar wind may modulate the corresponding turbulent heating, offering the ability to better understand how turbulence dissipates in the solar wind plasma. 

Recent work has begun to show a connection between switchbacks and turbulence. \cite{malaspina_inhomogeneous_2022} showed that broadband wave activity is observed to correlate with the occurrence of switchbacks and demonstrated that KAWs are preferentially observed during switchbacks, concluding that the heating effects of KAWs may also be correlated with switchbacks. \cite{dudok_de_wit_switchbacks_2020} showed that switchbacks are associated with a more well developed inertial range than surrounding quiescent regions, which suggests that these regions may more effectively transfer their turbulent energy to kinetic scales. Similarly, results by \cite{hernandez_impact_2021} showed enhanced cascade rates inside switchbacks, suggesting that switchbacks couple to the turbulent cascade in such a way as to enhance it. On the other hand, \cite{tenerani_evolution_2021} showed that switchback amplitude decreases faster than that of the turbulent fluctuations as a function of radial distance from the sun, suggesting that switchbacks may not be damping by coupling to the turbulent cascade. 

However, when studying the wave properties within switchbacks, care must be taken to account for the effects of solar wind anisotropy. The presence of a background magnetic field causes solar wind turbulence to become anisotropic \citep{goldstein_kinetic_2015,chen_interpreting_2010,matthaeus_scaling_1998}, with more power contained in fluctuations perpendicular to the background magnetic field than those parallel to the background field. It has been shown that the power spectrum measured by a single spacecraft is heavily dependent on the angle between the solar wind velocity in the spacecraft frame and the background magnetic field \citep{horbury_anisotropic_2008,podesta_dependence_2009,luo_observations_2010,forman_detailed_2011}, often referred to as $\theta_{vB}$. The field rotations and velocity spikes associated with switchbacks have the potential to locally modulate $\theta_{vB}$ and thereby modulate the effective measurements of wave power within these regions.

This paper uses observations from the Parker Solar Probe (PSP) \citep{fox_solar_2016} to investigate the relationship between turbulent wave power and switchback deflection, while controlling for $\theta_{vB}$. We measure integrated turbulent wave power in both the inertial and kinetic ranges to study the behavior of each regime. We find that, when controlling for the observational effects of the varying $\theta_{vB}$ within switchbacks, switchback deflection does not correlate to a significant increase in turbulent wave power, in either the inertial or kinetic ranges. We examine the effect of the angle $\theta_{vB}$ on the effective measurement of parallel versus perpendicular wave power and observe that changes in $\theta_{vB}$ due to switchback deflection leads to an apparent increase in turbulent wave power, due to a more effective sampling of perpendicular power. Once this effect is accounted for, no additional wave-power was observed within switchback regions, suggesting that switchbacks may not play a distinct dynamical role in solar wind turbulence and therefore would not be associated with turbulent dissipation and plasma energization via turbulent wave-particle interactions.  

Section \ref{sec:methods} of this paper outlines our methodology and approach to measuring the relationship between turbulent wave power and switchbakcs. Section \ref{sec:data} discusses the data selection and processing used in this study. Section \ref{sec:obs} presents our observations and Section \ref{sec:disc} discusses these results and their implications. Finally, Section \ref{sec:conc} summarizes and concludes the work presented in this paper.

\section{Methodology} \label{sec:methods}

In this study, we define switchback deflection as the angle $\alpha$ by which the background magnetic field deviates from the nominal Parker spiral. We adopt the definition of normalized switchback deflection angle $z$, as defined by Equation \ref{z_equation} \citep{dudok_de_wit_switchbacks_2020}. $z$ goes from 0 to 1 as the angle $\alpha$ goes from $0^{\circ}$ to $180^{\circ}$ for a background magnetic field oriented towards the sun.

\begin{equation}\label{z_equation}
    z = \frac{1}{2}(1-cos \alpha).
\end{equation}

$z=1$ (or $\alpha = 180^{\circ}$) corresponds to a fully reversed field. To preserve the meaning of z across changes in polarity when crossing the heliospheric current sheet, we let $z\rightarrow 1-z$ when the background field is pointing predominantly away from the sun. \cite{dudok_de_wit_switchbacks_2020} uses PSPs first orbital encounter to show that $z$ follows a power-law distribution, concluding that switchbacks have no typical deflection. However, they show that $z = 0.05$ serves as an effective threshold between switchback regions and the surrounding quiescent solar wind. Therefore, in this study, we adopt the definition that $z\ge0.05$ corresponds to a switchback and $z<0.05$ is a quiescent region.

To identify the transition to kinetic scales, we use a technique similar to that used in \cite{malaspina_inhomogeneous_2022} to identify a frequency $f_{break}$ (dark blue line in Figure \ref{fig:1hr_spectra_time_series}a,b) at which turbulent cascade transitions from inertial range to kinetic range. This approach compares the ratio of fluctuating perpendicular electric and magnetic fields to identify a transition to kinetic scales \citep{malaspina_inhomogeneous_2022,salem_identification_2012, bale_measurement_2005}. Eq. 48 in \cite{stasiewicz_small_2000} shows that this transition takes place near proton scales where $E_{\perp}/B_{\perp}$ rises above the local Alfvén speed (see Figure 3 in \cite{malaspina_inhomogeneous_2022}). In this work, $E_{\perp}/B_{\perp}$ spectra are binned into log-spaced frequency bins and $f_{break}$ is defined as the frequency at which the bin median rises above the local Alfvén speed.

To capture the power contained in kinetic scales, we integrate the  power spectral density  of the total magnetic field $B_{PSD}$ above $f_{break}$ as shown in Eq. \ref{kaw_power_eq}

\begin{equation}\label{kaw_power_eq}
    P_{HF} = \int_{f_{break}}^{f_{max}} B_{PSD}(f) \,df.
\end{equation}
    
\begin{figure*}[!htb]
\makebox[\textwidth][c]{\includegraphics[width=.3\textwidth]{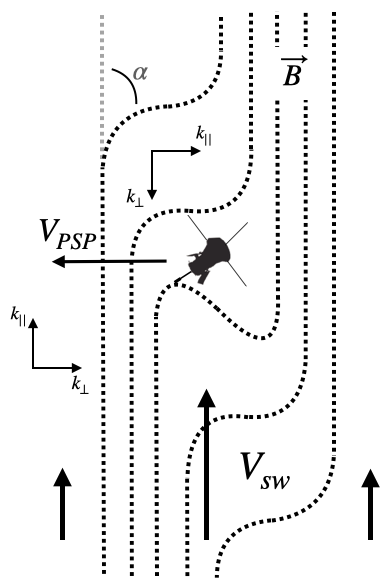}}
\caption{Cartoon graphic of PSP flying through a switchback structure. The combination of rotated fields and increased radial solar wind velocity within switchbacks advects the perpendicular power spectrum past the spacecraft (i.e. increases $\theta_{vB}$), increasing measured wave power.   
\label{fig:SB_graphic}}
\end{figure*}

We impose an upper integration bound $f_{max} = f_{break}+100Hz$ so that our power metric is independent of sample rate (and thereby frequency resolution) or effects of Doppler shifting.  To capture the corresponding low frequency power, we define Eq. $\ref{low_f_power_eq}$

\begin{equation}\label{low_f_power_eq}
    P_{LF} = \int_{f_{DC}}^{f_{break}} B_{PSD}(f) \,df ,
\end{equation}

where $f_{DC} \approx 10^{-1}$ Hz. To isolate the relationship between wave power and $z$, we must account for the effects of solar wind anisotropy. As illustrated in Figure \ref{fig:SB_graphic}, when the spacecraft enters a switchback, it enters a region of rotated field and increased radial solar wind velocity. This can have the effect of increasing $\theta_{vB}$, causing switchbacks to act as regions of the solar wind that allow for more effective sampling of perpendicular power. This will cause an apparent increase of observed wave power, when the switchback deflection angle is large. This effect must be disambiguated from increased wave power from enhanced turbulent dissipation, which we account for as described below.

We therefore choose to examine measured wave power as a function of $\theta_{vB}$ and look for additional power contributions from $z$ on top of those from $\theta_{vB}$ alone. Here we define $\theta_{vB}$ as $\text{tan}^{-1} (v_\perp/v_{||})$ where $v_\perp$ and $v_{||}$ are components of the spacecraft frame solar wind velocity perpendicular and parallel to the background magnetic field. $\theta_{vB} \approx 0$ means that the spacecraft is sampling parallel/antiparallel to the background field while $\theta_{vB} \approx 90$ corresponds to sampling perpendicular to the background field. We anticipate the relationship between power and $\theta_{vB}$ to approximate the geometric relationship $P = P_0(1-\text{cos}(\theta_{vB}))$, which will increase monotonically from $\theta_{vB}= 0^{\circ}$ to a maximum at $\theta_{vB}= 90^{\circ}$. Recent studies have examined the link between $theta_{vB}$ and solar wind turbulnence \citep{cuesta_isotropization_2022} and a detailed analysis by \cite{forman_detailed_2011} showed that this relationship can be described more rigorously by a wave vector spectrum in anisotropic "critical balance". However, our simplified form captures the lowest order behavior of the measured power being controlled by the geometry of the spacecraft trajectory with respect to the background field.

We take three approaches to viewing the relationship between turbulent wave power and switchback deflection. We first separate measurements of integrated wave power versus $\theta_{vB}$ by their existence inside our outside of switchback regions. For this study, we define switchback and quiescent regions as those that maintain $z\ge0.05$ and $z<0.05$, respectively, for at least 20s in the spacecraft frame. To reduce ambiguity, regions that do not maintain clear characteristics of being either a switchback or a quiescent region, as defined by this threshold on z, are excluded. This approach is expected to most clearly demonstrate differences in wave power between the two types of regions. We then examine the same relationship, but without selecting for clearly defined switchback/quiescent regions. Rather, we include all data points and observe the effect that the corresponding point-by-point value of z has on the measured wave power. Since integrating the power spectrum hides information about about its shape, we take a third approach that focuses on a comparison of the full power spectrum (rather than the integrated spectrum) when sorted by thresholds on both $\theta_{vB}$ and $z$.

\section{Data Set and Processing} \label{sec:data}

This study uses data from Parker Solar Probe, which launched on August 12, 2018 \citep{fox_solar_2016}. We use data from the 1st, 6th, 9th, and 11th near-Sun encounters during perihelion. Encounter 1 was chosen because it provides data at the greatest heliocentric distance during perihelion. The perihelion passes of encounters 6, 9, and 11 that are used here, offer three independent orbital groupings interior to encounter 1, so that possible radial trends can be illuminated. Data from the 1st encounter is taken between November 5th and 6th of 2018 with radial coverage of $\sim$ 35.5 - 36.5 $R_S$. Data from the 6th encounter is taken between September 26th and 28th of 2020 with radial coverage of $\sim$ 20 - 26 $R_S$. Data from the 9th encounter is taken between  August 9th and 10th of 2021 with radial coverage of $\sim$ 16 - 19 $R_S$. Data from the 11th encounter is taken between  February 25th and 26th of 2022 with radial coverage of $\sim$ 13 - 16 $R_S$. 

Electric and magnetic field measurements are obtained from the FIELDS instrument suite \citep{bale_fields_2016,malaspina_digital_2016,pulupa_solar_2017}. FIELDS measures DC-coupled electric fields in the plane of the spacecraft heat shield, three-axis DC-coupled magnetic fields from the fluxgate magnetometer (FGM), and three-axis AC-coupled magnetic fields from the search coil magnetometer (SCM). However, after encounter 1, one of the three orthogonal components of the SCM (the u-component in uvw sensor coordinates) is not reliable. For these times we approximate the missing component as the mean value of the remaining two. This loses phase information but reasonably approximates total power since the SCM is mounted symmetrically \citep[see Figure 10 in][]{bale_fields_2016} with respect to a predominantly radial background field. To convert electric potential to electric field, we assume an effective antenna length of 2m. The FGM data are sampled continuously at a cadence of ~293 samples per second (sps) for all encounters used in this study. The SCM and electric field data are sampled at cadences of $\sim$293 sps for encounter 1, $\sim$586 sps for encounter 6, $\sim$2343 sps for encounter 9, and $\sim$586 sps for encounter 11. We determine the background magnetic field using a 10s running window median filter of each component of the FGM data. Total plasma density is determined from FIELDS quasi-thermal noise measurements \citep{moncuquet_first_2020}.

Time series FIELDS data are rotated into field aligned coordinates (FAC) using the background field as defined above. FIELDS data are processed into spectra using windowed fast Fourier transforms. The window width is fixed at $\sim$6.9s and a Hanning window function is used with 50$\%$ overlap between neighboring windows. 

Proton distribution moments are obtained from Solar Wind Electrons Alphas and Protons (SWEAP)\citep{kasper_solar_2016, case_solar_2020, whittlesey_solar_2020,livi_solar_2022}. Proton velocity and core temperature are determined using data from the Solar Probe Cup (SPC) during encounter 1 and the ion Solar Probe Analyzer (SPANi) during encounters 6, 9, and 11, when the SPANi field of view is able to capture the proton core distribution.

\section{Observations} \label{sec:obs}

\begin{figure*}[!htb]
\makebox[\textwidth][c]{\includegraphics[width=.8\textwidth]{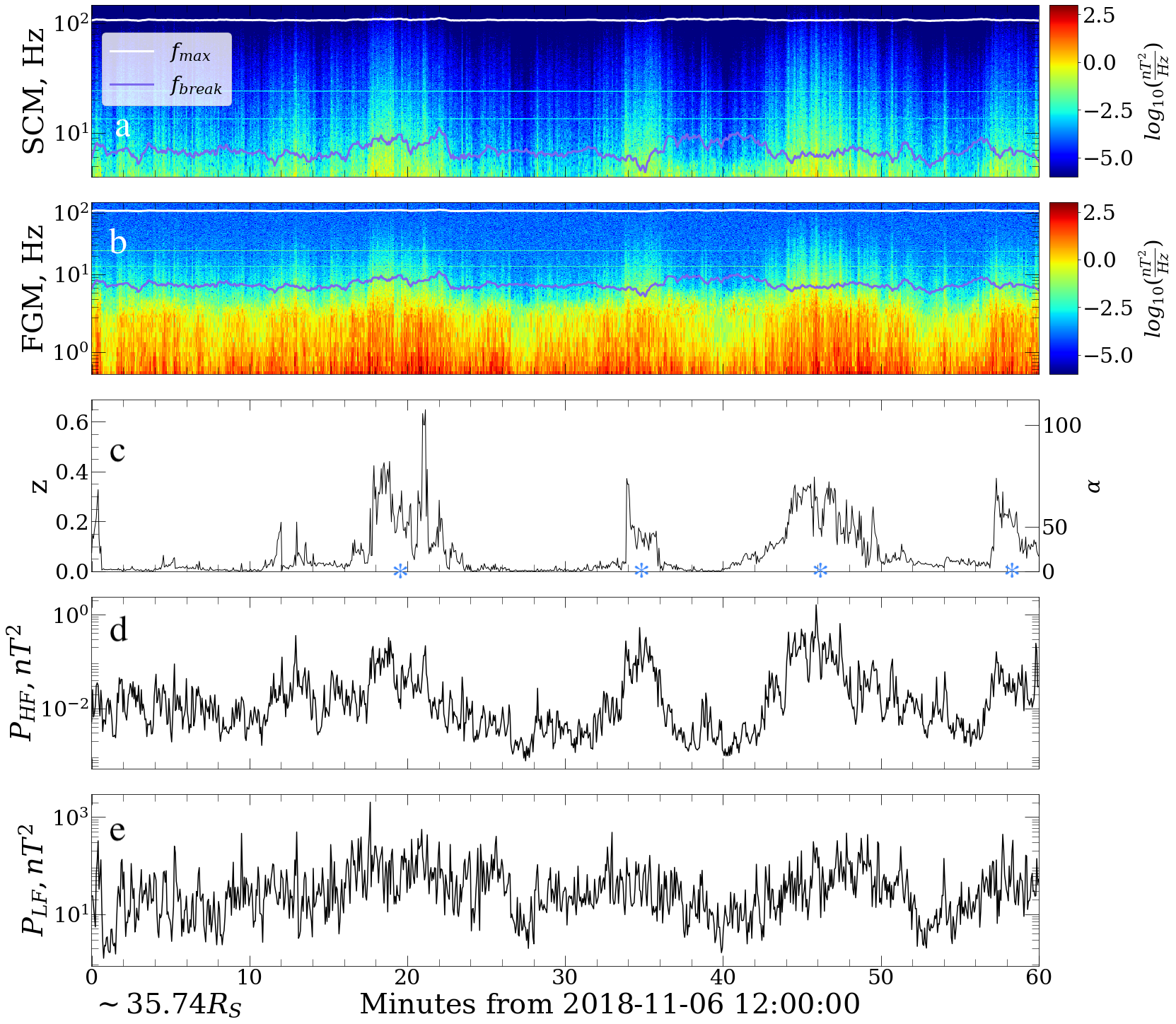}}
\caption{One hour example from Encounter 1 showing magnetic power spectra from (a) the SCM and (b) the FGM. The dark line shows the kinetic scale break frequency $f_{break}$ (see text for discussion) and the white line shows the corresponding upper integration bound $f_{max}$. (c) Normalized deflection angle $z$. Blue asterisks mark switchback patches (d) Kinetic scale wave power, $P_{HF}$. (e) Intertial range wave power, $P_{LF}$. 
\label{fig:1hr_spectra_time_series}}
\end{figure*}

Figure \ref{fig:1hr_spectra_time_series} shows a one hour example of plasma wave activity from encounter 1 at 35.74 $R_S$. Figure \ref{fig:1hr_spectra_time_series}a shows the magnetic power spectral density (PSD) from the SCM (4Hz to $\sim$150Hz). Figure \ref{fig:1hr_spectra_time_series}b shows the magnetic PSD from the FGM (DC to $\sim$150Hz ). $f_{break}$ is shown as a blue line and $f_{max}$ is shown as a white line in Figure \ref{fig:1hr_spectra_time_series}a,b. Figure \ref{fig:1hr_spectra_time_series}c shows the normalized switchback deflection angle $z$. Examples of switchback regions are marked with blue stars in Figure \ref{fig:1hr_spectra_time_series}c. An increase in broadband wave power in panels (a) and (b) can be seen during regions of increased z, suggesting a correlation between switchback deflection and turbulent wave power. Integrated high and low frequency power,  $P_{HF}$ and $P_{LF}$ are shown in Figure \ref{fig:1hr_spectra_time_series}d and \ref{fig:1hr_spectra_time_series}e, respectively. As z increases, an increase in both the high and low frequency regimes can be seen.

\begin{figure*}[!htb]
\makebox[\textwidth][c]{\includegraphics[width=1\textwidth]{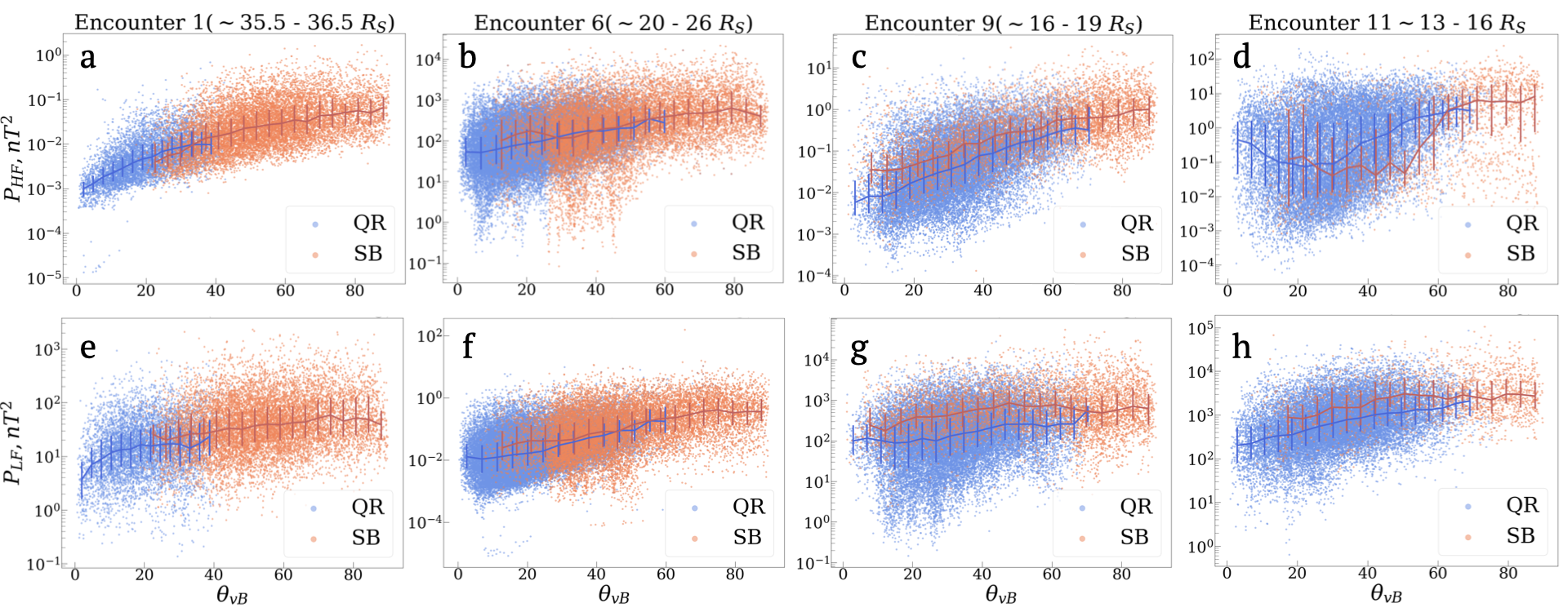}}
\caption{ (top row) Kinetic scale wave power $P_{HF}$ vs. $\theta_{vB}$ and (bottom row) inertial range wave power $P_{LF}$ vs. $\theta_{vB}$ with columns showing separate encounters. Orange and blue points represent those associated with switchback and quiescent regions, respectively. Each distribution (switchback and quiescent) is separated into 20 bins along the $\theta_{vB}$ axis and bin medians are shown with solid lines. Bins with less than 50 counts are excluded. Error bars represent 1st and 3rd bin quartiles.
\label{fig:p_thetavb_qrsb_sorted}}
\end{figure*}

Figure \ref{fig:p_thetavb_qrsb_sorted} shows integrated turbulent wave power as a function of $\theta_{vB}$. The top row shows $P_{HF}$ vs. $\theta_{vB}$, the bottom row shows $P_{LF}$ vs. $\theta_{vB}$, and each column is a separate orbital encounter. As expected, wave power increases as $\theta_{vB}$ tends towards $90^{\circ}$. Orange and blue dots represent points associated with switchback and quiescent regions, respectively. Corresponding colored lined represent median values of the binned data with error bars representing the 1st and 3rd quartiles. One takeaway from Figure \ref{fig:p_thetavb_qrsb_sorted} is that switchback regions (orange) tend to be associated with larger values of $\theta_{vB}$. Most importantly, we note that quiescent and switchback regions form one continuous distribution. That is, switchback regions do not appear to show extra wave power in addition to the increase in power caused by measurement geometry $\theta_{vB}$. In particular, for the range of values of $\theta_{vB}$ where quiescent and switchback regions overlap, there is no clear difference in power. Figure \ref{fig:p_thetavb_qrsb_sorted}c
shows the clearest suggestion of increased kinetic scale wave power inside switchbacks, as the orange median line for the switchback distribution is shifted above that of the quiescent regions, but the difference is not significant compared to the spread in power. Similarly, panels g and h of Figure \ref{fig:p_thetavb_qrsb_sorted} suggest a possible increase in inertial range wave power inside switchbacks, but again the spread is too large to draw clear conclusions. 

\begin{figure*}[!htb]
\makebox[\textwidth][c]{\includegraphics[width=.8\textwidth]{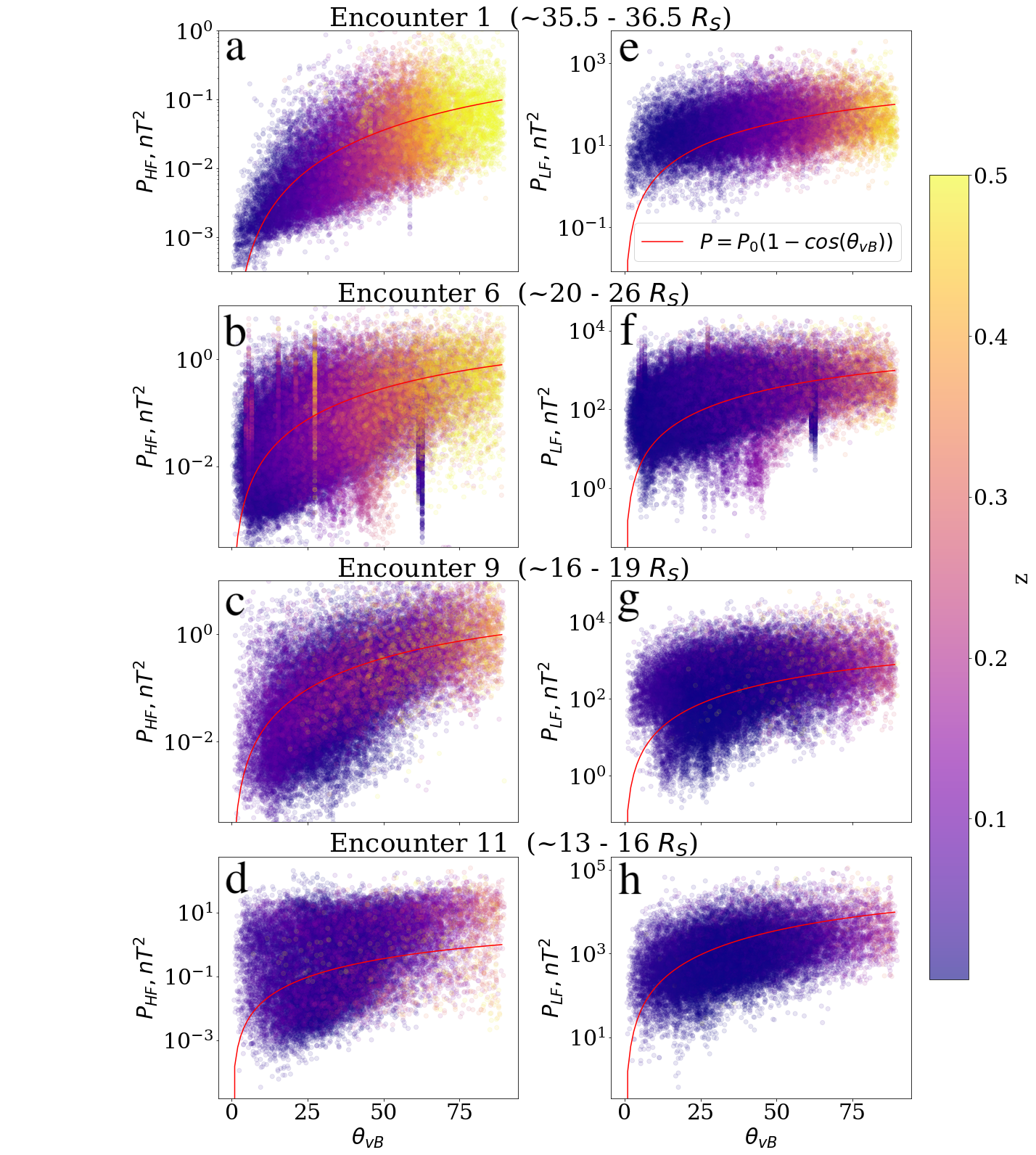}}
\caption{Kinetic scale (inertial range) integrated turbulent wave power $P_{HF}$ ($P_{LF}$) vs. $\theta_{vB}$ for encounters 1,6,9,and 11 in panels a-d (e-h). The value of $z$ for each data point is represented by the color bar. Curves of the form $P = P_0(1-\text{cos}(\theta_{vB}))$ are shown in red with $P_0$ qualitatively scaled to fit each panel.     
\label{fig:P_thetavB_continuous}}
\end{figure*}

Figure \ref{fig:P_thetavB_continuous} again shows $P_{HF}$ vs. $\theta_{vB}$ (a-d) and $P_{LF}$ vs. $\theta_{vB}$ (e-h) for each encounter. Unlike Figure \ref{fig:p_thetavb_qrsb_sorted}, Figure \ref{fig:P_thetavB_continuous} does not sort the data by its existence within or without a switchback. Rather, the associated value of $z$ for each point is represented by the color bar. Each panel of Figure \ref{fig:P_thetavB_continuous} also shows a red curve of the form $P = P_0(1-\text{cos}(\theta_{vB}))$ with $P_0$ scaled to fit the each distribution. The data follow this geometric relationship, especially at kinetic scales (a-e), where anisotropy is greatest. Again, the data follow a single distribution and power does not show any dependence on $z$. For any given value of $\theta_{vB}$, power and $z$ show no dependence on one another. We also see again that $z$ tends to increase with increasing $\theta_{vB}$, more clearly demonstrating the correlation between $z$ and $\theta_{vB}$ that was suggested by Figure \ref{fig:p_thetavb_qrsb_sorted}.

We also note a trend in the distribution of $z$ with radial distance that shows a greater number of large switchback deflection angles ($z \approx .5$)($z>.5$ is not statistically well represented in this study) at larger radial distances form the sun.

\begin{figure*}[htb]
\makebox[\textwidth][c]{\includegraphics[width=.4\columnwidth]{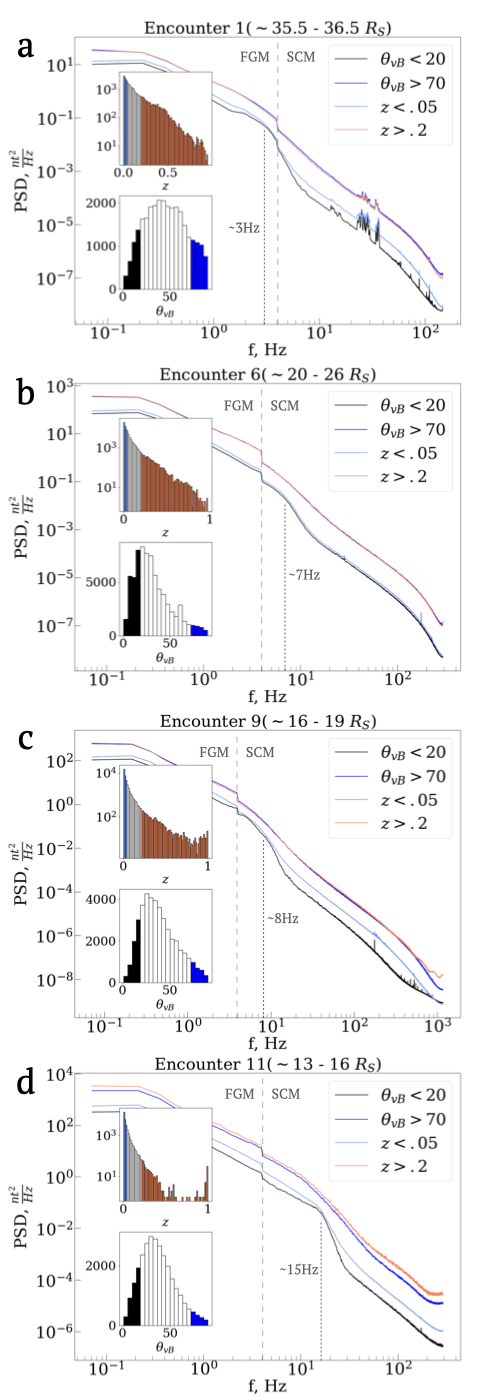}}
\caption{Average power spectra for times with $\theta_{vB}<20$ (black), $\theta_{vB}>20$ (dark blue), $z<.05$ (light blue), and $z>.2$ (orange). Sampling statistics are shown in the inset distributions on the left side of each plot. A vertical dashed line is placed at 4Hz on each plot to mark the transition between spectra calculated using the FGM and the SCM. A separate vertical line is placed at the frequency corresponding to the center of the bump in the quasi-parallel ($\theta_{vB}<20$) spectra, which is possibly associated with ion cyclotron wave power.        
\label{fig:PSD_sorted}}
\end{figure*}

Figure \ref{fig:PSD_sorted} shows average power spectra for quasi-parallel ($\theta_{vB}<20$)(black) and quasi-perpendicular ($\theta_{vB}>70$)(dark blue) sampling of the turbulent wave field as well as for times with $z < .05$(light blue) and $z > .2$(orange). Encounters 1, 6, 9, and 11 are shown in panels a, b, c, d of Figure \ref{fig:PSD_sorted}, respectively. A vertical dashed line is placed at 4Hz on each panel to denote the transition between spectra calculated using the FGM and the SCM. Inset in each panel are the distributions of $z$ and $\theta_{vB}$ for each encounter with color coding showing which parts of the distribution contribute to each average spectra. Again, we note an increasing proportion of larger values of $z$ with increasing radial distance.

 A difference in shape can be seen between the quasi-parallel and quasi-perpendicular spectra. In particular, the presence of ion cyclotron power can be seen in the $\theta_{vB}<20$ spectrum near 3, 7, 8, and 15 Hz for encounters 1, 6, 9, and 11, respectively. The $\theta_{vB}>70$, or quasi-perpendicular spectrum, contains more power, especially at higher frequencies.  Most importantly, the average spectra for large values of $z$ converges with the quasi-perpendicular spectrum. Conversely, when $z$ is low, the measured power spectrum tends towards the quasi-parallel spectrum.

\section{Discussion} \label{sec:disc}

This study addresses the outstanding question of the relationship between both the generation and evolution of switchbacks and turbulent dissipation by evaluating the relationship between the amplitude of local turbulent fluctuations and switchback deflection angle $z$. To separate out the effect of the anisotropic distribution of wave power in the solar wind, we do so while controlling for the the angle between spacecraft trajectory and the background magnetic field $\theta_{vB}$.

We show in Figures \ref{fig:p_thetavb_qrsb_sorted} and \ref{fig:P_thetavB_continuous} that as $\theta_{vB}$ increases from $0^{\circ} - 90^{\circ}$, integrated wave power increases by 2 to 3 orders of magnitude within kinetic scales and roughly 1 order of magnitude in the inertial range. This is a larger anisotropy than that measurd by \cite{podesta_dependence_2009} at 1AU who showed a power ratio ($\frac{P_\perp}{P_{||}}$) of a few in the inertial range and $\sim$15 at kinetic scales. However, our results are more consistent with theoretical estimates by \cite{chen_interpreting_2010} that predict a power ratio of $\sim$5 in the inertial range and $\sim$400 near $k_\perp\rho_i=1$, where $\rho_i$ is the ion gyroradius.  

If switchback and quiescent regions contained differing amounts of inherent wave power, they would create two separate distributions of power versus $\theta_{vB}$, for the same range of $\theta_{vB}$. However, Figure \ref{fig:p_thetavb_qrsb_sorted} demonstrates that these distinct regions form one single distribution, suggesting that there is no difference in the amount of wave power contained within them. Rather, switchback (quiescent) regions tend to be associated with larger (smaller) values of $\theta_{vB}$, which causes an apparent increase in wave power.  

It may be expected that switchbacks would correspond to increased large scale power, since they themselves are large scale structures. However we do not see clear evidence of an increased distribution of $P_{LF}$ vs $\theta_{vB}$ in Figure \ref{fig:p_thetavb_qrsb_sorted}(e-h). Only a slight suggestion of such an increase in panels g and h. This ambiguity may be related to the fact that 
we are not examining a sufficiently low wave number (frequency) portion of the spectra.

Figure \ref{fig:P_thetavB_continuous} offers a representation of the data that allows us to examine the effect that $z$ has on the data point-by-point, regardless of whether or not the spacecraft is inside or outside of a clearly defined switchback region. Again, we see that power is only modulated by measurement geometry and the local value of $z$ does not have any additional effect.  

In comparison to Figures \ref{fig:p_thetavb_qrsb_sorted} and \ref{fig:P_thetavB_continuous}, which solely examine integrated wave power, Figure \ref{fig:PSD_sorted} employs spectral shape to further demonstrate that the modulation of measured power by $z$ is due to a modulation of the entire spectrum towards the perpendicular and parallel limits of sampling. Similar to \cite{bale_highly_2019} and \cite{bowen_mediation_2024}, we see the presence of ion cyclotron power in the low $z$ (low $\theta_{vB}$) spectrum.

These observations show that measurements of wave power inside switchbacks are mainly affected by the pre-existing solar wind anisotropy, and do not appear to be affected by the switchbacks themselves. These swicthback regions of rotated magnetic fields and increased radial bulk flow tend to increase the quantity $\theta_{vB}$ towards $90^{\circ}$, which causes more effective measurement of perpendicular power as described earlier. We therefore conclude that this apparent increase in  wave power within switchbacks is not due to an increase in inherent wave power in the plasma but, rather, is a measurement effect expected by the change in sampling of the anisotropic solar wind turbulent fluctuations.  

The apparent absence of increased wave power associated with switchbacks suggests that wave-mediated turbulent dissipation does not appear to be enhanced within these structures, and therefore switchbacks would not contribute to the turbulent energy cascade and associated energization via this pathway of energy conversion. However, further study is needed in order to obtain a more coherent understanding of the properties of turbulent energy transfer within switchbacks and quiescent regions in the inner heliosphere. Mechanisms of intermittent dissipation such as magnetic reconnection, which has been observed to occur in current sheets both in switchbacks and quiescent regions \citep{froment_direct_2021}, also contribute to turbulent dissipation, leading to particle heating and acceleration, and their relative importance in the turbulent dynamics has yet to be determined. 

It is expected that measurements of the spatial distribution of fluctuations within switchbacks made by a single spacecraft will be similarly affected by the close relationship between $z$ and $\theta_{vB}$. This biasing towards perpendicular sampling within switchbacks may affect efforts to understand the difference in turbulent characteristics between switchbacks and quiescent regions. 
Therefore, further studies on the properties of plasma turbulence within switchbacks, while controlling for $\theta_{vB}$ are needed to help determine if the relationship between turbulence and switchbacks could be mediated via other energy conversion pathways.
Moreover, future multi-spacecraft missions such as HelioSwarm and Plasma Observatory will be able to take advantage of simultaneous multi-point, multi-scale measurements of the solar wind, giving us a more complete picture of the nature of solar wind turbulence, examining the role of anisotropies and cross-scale energy trasnfer.

An additional observation presented in this study is that of the evolution of the distribution of $z$ with radial distance from the sun. Figures \ref{fig:p_thetavb_qrsb_sorted} and \ref{fig:P_thetavB_continuous} both show an increasing proportion of points associated with larger values of $z$ with increasing radial distance. This agrees with observations by \citep{pecora_magnetic_2022} that switchback counts, when switchbacks are defined as full reversals of the background field, increase with increasing radial distance from the sun. Figure \ref{fig:PSD_sorted} more clearly shows the evolution of the shape distribution of $z$ with radial distance. The distribution starts at closer radial distances fro the sun with a polynomial shape with low counts at large values of z and evolves with radial distance to fill in larger values of $z$ and take on a more exponential shape, like that reported by \cite{dudok_de_wit_switchbacks_2020} during PSP's 1st encounter. This increase in switchback deflection angle with radial distance may be evidence of the existence of an in-situ process (i.e. shearing) that amplifies the field reversal of switchbacks.

\section{Conclusion} \label{sec:conc}

This study uses the integrated magnetic power spectrum to characterize the relationship between turbulent wave power and switchback deflection angle. We show that switchback parameter $z$ strongly controls $\theta_{vB}$ and, thereby, the sampling of the anisotropic turbulent wave field of the solar wind. We conclude that the apparent increase in turbulent wave power within switchbacks is solely due to a more effective sampling of perpendicular power. Our results are consistent with previous spacecraft measurements of solar wind anisotropy as a function of $\theta_{vB}$. These observations suggest that switchbacks do not play a dynamical role in solar wind turbulence and, therefore, should not locally modify the cascade in such a way as to increase heating. This, however, does not rule out other dissipation mechanisms for switchbacks. Which pathways of energy conversion are accessed by switchbacks, if any, and how switchbacks contribute overall to solar wind heating remains an open question.

\section*{Acknowledgments}

Parker Solar Probe was designed, was built, and is now operated by the Johns Hopkins Applied Physics Laboratory as part of NASA’s Living with a Star (LWS) program (contract NNN06AA01C). Support from the LWS management and technical team has played a critical role in the success of the Parker Solar Probe mission.

\section*{Data Availability}
The data used in this study are available at the NASA Space Physics Data Facility (SPDF): https://spdf.gsfc.nasa.gov.

\bibliography{references}{}
\bibliographystyle{aasjournal}



\end{document}